\documentclass[12pt]{article}
\usepackage[margin=1in]{geometry}
\usepackage{amsmath,amssymb,draft}
\usepackage{graphicx}
\usepackage{dcolumn}
\usepackage{bm,tikz-cd}
\makeatletter
\newenvironment{widetext}{\par\noindent}{\par}
\makeatother

\def\ma{\mathcal}
\def\ie{\begin{equation}\begin{aligned}}
\def\fe{\end{aligned}\end{equation}}

\begin{document}

\title{Off-shell recursion for all-loop planar integrands in Yang-Mills theory}
\author{Yi-Xiao Tao}
\email{taoyx21@mails.tsinghua.edu.cn}
\address{Department of Mathematical Sciences, Tsinghua University, Beijing 100084, China}
\date{}

\begin{abstract}
In this paper, we develop in detail the off-shell recursion for planar loop integrands in Yang-Mills theory. Starting from the classical equations of motion solved with the perturbiner method, we derive an exact transfer-matrix representation of the pure-gluon sector. We then include the ghost contributions to the loop kernels based on \cite{Tao:2025fch}. Finally, as an example, we work out the two-loop recursion in detail and conclude a general recursion strategy for two-loop planar integrands whose external legs are gluons.
\end{abstract}

\section{Introduction}
In the past several years, many amplitude relations at the 1-loop level have been discovered \cite{Cao:2024olg,Zhou:2021kzv,Zhou:2022djx,Cao:2025ygu,Du:2025rty,Du:2025yxz,Xie:2025utp,Chen:2023bji}. Based on these successes, a natural thought is to study amplitude relations at the higher-loop level. However, it is very difficult to construct higher-loop integrands in an organized way, which will be convenient for us to find amplitude relations. The most popular way now is still using Feynman rules and summing over all Feynman diagrams, which mainly depends on the programme and is difficult to help us explore integrand structures. Recently, a more formal way based on the ``surfaceology" has been proposed \cite{Arkani-Hamed:2023lbd,Arkani-Hamed:2023mvg,Arkani-Hamed:2024tzl,Arkani-Hamed:2024pzc,Cao:2025mlt}, which is also an on-shell method. This recursive construction may shed light on the amplitude relations at the higher-loop level, but it still needs to be developed.

An alternative method for constructing higher-loop integrands recursively is proposed in \cite{Tao:2025fch,Tao:2025pnt}. This method begins with the off-shell multi-particle currents, which come from the classical equation of motion, and is therefore an off-shell method. The basic idea for this off-shell method is sewing some off-shell tree-level objects to construct loops in an organized way, and the amplitude relations for the off-shell tree-level objects \cite{Chen:2023bji,Tao:2022nqc,Tao:2024vcz,Wu:2021exa,Tao:2023yxy} will be generalized to the loop-level through some sewing methods \cite{Gomez:2022dzk,Chattopadhyay:2021udc}. This method is based solely on the classical equation of motion; therefore, it has a wide range of applications. This method is actually a generalization of the Berends-Giele recursion \cite{Berends:1987me,Berends:1988zn} which has broad applications \cite{Mafra:2010jq,Mafra:2015vca,Mafra:2016ltu,Lee:2022aiu,Armstrong:2022mfr,Cho:2021nim,Cho:2022faq,Cho:2023kux,Gomez:2021shh,Tao:2023wls,Chattopadhyay:2024kdq}.

In this paper we focus on the Yang-Mills (YM) theory because of its significance for amplitude relations. In \cite{Tao:2025fch}, the YM theory was discussed briefly as part of a general framework for $\ell$-loop planar integrands in colored theories, and a set of two-loop gluon examples was checked against Feynman rules. In the present paper we add three new ingredients and justify them in detail:
\begin{enumerate}
\item We show that the pure-gluon sector admits a transfer-matrix (block-matrix) formulation. More precisely, we prove by induction that every comb component of a multi-particle gluon current is obtained by multiplying $2\times2$ block matrices, one per vertex, and that the one-loop kernel factorizes into a chain of such matrices together with a contact seed matrix. We give the precise index and contraction conventions of the factorization, and we explain why this representation makes the sewing step local: a given leg appears only in the two matrices adjacent to its position in the chain, so that the sewing of a comb component into a lower-loop kernel only modifies a finite, leg-independent part of the integrand.
\item We extend the recursion to the complete YM theory by including the Faddeev-Popov ghosts. We derive the ghost multi-particle currents and their comb components from the equations of motion, state explicitly the Grassmann parity and sign conventions used at every step of the recursion, and translate them into a definite set of sewing rules for ghost legs. This closes the gap between the pure-gluon construction of \cite{Tao:2025fch} and the full YM theory.
\item We work out the two-loop recursion in detail. For the two-point two-loop case we enumerate the ordered sets and $k$-divisions of the recursion, identify the one-loop kernels that are actually needed, and explain, through the graph factors and the implicit symmetries of the construction, why the other one-loop kernels do not contribute. This leads to a general recursion strategy for two-loop planar integrands whose open legs are all gluonic.
\end{enumerate}
Besides these technical results, we take the opportunity to make the presentation self-contained: all objects that enter the recursion, such as the comb and contact components, the loop kernels, and the graph factors, are defined and derived in the main text, and the derivation of the central formulas is given step by step. This is useful both for readers who are not familiar with \cite{Tao:2025fch} and for making the logical status of each ingredient of the recursion transparent. Throughout the paper we work with the planar part of the loop integrand, which is the part relevant for the single-trace partial amplitudes in the color decomposition, and we use the Feynman gauge for the off-shell currents; the dependence of intermediate objects on this gauge choice is discussed in section \ref{sec2}.

This paper is organized as follows. In section \ref{sec2} we introduce the basic concepts, including the multi-particle currents of the YM theory, their comb and contact components, the one-loop kernels, the graph factors, and the recursive formula for $\ell$-loop planar integrands. In section \ref{sec3}, we focus on the pure-gluon sector and derive the transfer-matrix formalism: we prove that the comb components satisfy a transfer-matrix recursion, show that the one-loop kernel can be written as a product of vertex matrices, and discuss the local properties of the sewing procedure that follow from this representation. In section \ref{sec4}, we include the ghosts. We derive the ghost multi-particle currents and their comb components, state the Grassmann conventions used in the recursion, present the corresponding one-loop kernels, and write down the complete loop-kernel recursion involving ghost fields. In section \ref{sec5}, we consider the two-loop planar integrand recursion as an example, focusing on the two-point two-loop case, and we conclude a recursion strategy for general two-loop planar integrands. Section \ref{sec6} contains our conclusions and outlook. Technical material on the graph factors and on the bi-adjoint scalar recursion, which is used as a warm-up for the YM case, is collected in the appendices.

\section{Basic concepts}\label{sec2}
\subsection{Multi-particle currents}
In this section we review the objects that are needed in the rest of the paper: the multi-particle currents of the Yang-Mills (YM) theory, their comb and contact components, and the loop kernels. The reader is referred to \cite{Tao:2025fch,Tao:2025pnt} for a fuller discussion of the general framework and of the bi-adjoint scalar realization of the recursion; the graph factors used in the loop-kernel recursion are defined systematically in appendix \ref{app:graph}.

We work with the color-ordered sector of the theory, i.e.\ with the part of the integrand that contributes to the single-trace partial amplitudes. All fields are Lie-algebra valued, $A_\mu=A_\mu^a T^a$, with generators normalized by $\mathrm{Tr}\,(T^aT^b)=\delta^{ab}$. For a word $P=p_1p_2\cdots p_m$ we write $k_P=k_{p_1}+\cdots+k_{p_m}$, $s_P=k_P^2$, and $T^{a_P}=T^{a_{p_1}}\cdots T^{a_{p_m}}$. A sum over ``deconcatenations'' $P=XY$ (respectively $P=XYZ$) means a sum over all ways of splitting the word $P$ into two (three) non-empty consecutive subwords $X,Y$ (respectively $X,Y,Z$). All external momenta are kept off-shell, $k_{p_i}^2\neq 0$, until the last step of the construction, where the on-shell limits of the external legs are taken; the only exceptions are the single-particle ``fields'' $A_{p\mu}$, $b_p$, $c_p$ that appear as the seeds of the recursion, which are functions of the momenta and polarizations and never acquire a propagator factor $1/s_P$ on their own.

For the Yang-Mills theory we use the following Lagrangian in the Feynman gauge \cite{Gomez:2022dzk}:
\ie
\ma{L}=\text{Tr}\,\left( -\frac{1}{4}F_{\mu\nu}F^{\mu\nu}-\frac{1}{2}(\partial_{\mu}A^{\mu})^2+\partial^{\mu}b(\partial_{\mu}c-i[A_{\mu},c])\right).
\fe
The last term is the standard Faddeev-Popov ghost action in the Feynman gauge, written with $b$ denoting the antighost and $c$ the ghost; both are Grassmann-odd. The multi-particle currents are defined through the perturbiner expansion \cite{Selivanov:1997aq,Selivanov:1998hn,Rosly:1997ap,Rosly:1998vm,Mizera:2018jbh,Lopez-Arcos:2019hvg}
\ie
\ma{A}_\mu(x)&=\sum_P A_{P\mu}\,e^{ik_P\cdot x}\,T^{a_P},\\
b(x)&=\sum_P b_P\,e^{ik_P\cdot x}\,T^{a_P},\qquad c(x)=\sum_P c_P\,e^{ik_P\cdot x}\,T^{a_P},
\fe
where the sums run over all words $P$. The coefficients $A_{P\mu},b_P,c_P$ are the \textit{multi-particle currents}. They solve the classical equations of motion order by order in the amplitude: inserting the ansatz above into the equation of motion and comparing the coefficient of $e^{ik_P\cdot x}$ gives an algebraic equation in which $k_P^2$ multiplies the current of word $P$ on the left-hand side, while the right-hand side is built from products of currents of proper subwords of $P$. The structure of the right-hand side is dictated by the interaction vertices of the Lagrangian, and the deconcatenation sums enumerate exactly the ways in which the subwords $X,Y,Z$ can join into $P$. For example, the coefficient of $e^{ik_P\cdot x}$ receives contributions from the three-gluon vertex through the two-deconcatenations $P=XY$, and from the four-gluon vertex through the three-deconcatenations $P=XYZ$, for which $k_X+k_Y+k_Z=k_P$. When we define the comb and contact components below, we follow the convention of \cite{Tao:2025fch} and read each three-deconcatenation as a pair of nested two-deconcatenations, $P=XW$ with $W=YZ$; this makes the recursive definition of the components uniform and is the reason why the quartic-vertex terms of \eqref{gluoncurrent} are organized in the way shown below. Explicitly, the gluon recursion reads
\ie\label{gluoncurrent}
k_{P}^{2}A_{P\mu}&=\sum_{P=XY}(k_{X\mu}-k_{Y\mu})(A_{X}\cdot A_{Y})-A_{X\mu}(k_{X}\cdot A_{Y})\\
&-A_{X\mu}(k_{P}\cdot A_{Y})+A_{Y\mu}(k_{Y}\cdot A_{X})+A_{Y\mu}(k_{P}\cdot A_{X})\\
&+\sum_{P=XYZ}2A_{Y\mu}(A_{X}\cdot A_{Z})-A_{X\mu}(A_{Y}\cdot A_{Z})\\
&-A_{Z\mu}(A_{X}\cdot A_{Y})+\sum_{P=XY}k_{Y\mu}b_{Y}c_{X}-k_{X\mu}b_{X}c_{Y}
\fe
which is the recursion used throughout this paper. The dot products denote the Lorentz contraction $A_X\cdot A_Y=A_{X\mu}A_Y^\mu$. The first two lines arise from the three-gluon vertex, the next two from the four-gluon vertex, and the last line from the gluon-ghost-antighost vertex; the ghost terms are discussed in detail in section \ref{sec4}. Note that the recursion is exact for the off-shell currents, in the sense that no on-shell or gauge-truncation condition has been imposed on any of the legs. This is the key property that makes the construction of loop integrands possible: every internal line of a later loop kernel will be produced by sewing two off-shell legs of such currents.

\subsection{Comb and contact components}
The recursion \eqref{gluoncurrent} involves a sum over all deconcatenations, and the full multi-particle current $A_{P\mu}$ is the sum of many different branchings of the word $P$. For the construction of loop integrands it is very useful to single out one special class of branchings, the \textit{comb component}. It is defined by the following requirement on the recursive decomposition: whenever the current of word $P$ is expressed through \eqref{gluoncurrent}, we keep only the contribution in which the \textit{last} letter $P_m$ is separated from the rest, i.e.\ $Y=P_m$ is a single letter in the two-deconcatenation $P=XY$, and, for the quartic-vertex part (written as a pair of nested two-deconcatenations, as explained above), both $Y$ and $Z$ are single letters, i.e.\ $P=XYZ$ with $Y=P_{m-1}$, $Z=P_m$. In other words, at every step of the recursion a new single letter is attached to the current of the word that contains all preceding letters; the resulting tree is a comb (a chain), hence the name. The same definition is applied recursively, starting from the single-letter currents $A_{i\mu}$.

To illustrate the definition, consider the comb component of $A_{l12\mu}$, where we denote the first off-shell leg by $l$ for later convenience:
\ie
&k_{l12}^2A_{l12\mu}^{\text{comb}}=(k_{l1\mu}-k_{2\mu})(A_{l1}\cdot A_{2})-A_{l1\mu}(k_{l1}\cdot A_{2})\\
&-A_{l1\mu}(k_{l12}\cdot A_{2})+A_{2\mu}(k_{2}\cdot A_{l1})+A_{2\mu}(k_{l12}\cdot A_{l1})\\
&+2A_{1\mu}(A_{l}\cdot A_{2})-A_{l\mu}(A_{1}\cdot A_{2})-A_{2\mu}(A_{l}\cdot A_{1})
\fe
Note that $A_{l1\mu}=A^{\rm comb}_{l1\mu}$ because a two-letter word admits only the splitting $(l,1)$. The first two lines of the example come from the three-gluon vertex, and the last line from the quartic-vertex part; the latter has the structure $2A_{1\mu}(A_l\cdot A_2)-A_{l\mu}(A_1\cdot A_2)-A_{2\mu}(A_l\cdot A_1)$, in which the index $\mu$ can sit on any of the three single letters, corresponding to the three inequivalent ways of routing the off-shell momentum through the quartic vertex.

\subsection{One-loop kernels and the contact component}
Then we can obtain the $m$-way 1-loop bare kernel from the comb component:
\ie\label{bk}
I^{\rm bare\ kernel}_{1,m}=k_{l12\cdots m}^2(A^{\rm comb}_{l12\cdots m}+A^{\rm contact}_{lm12\cdots m-1})\cdot \epsilon\bigg|_{\epsilon_{l\mu}\epsilon_{\nu}\to\frac{\eta_{\mu\nu}}{l_1^2}}
\fe
where the contact component $A^{\rm contact}_{P\mu}$ is used to maintain the cyclic invariance of the 1-loop kernel. The definition of a contact term $A_{P\mu}^{\rm contact}$ is that we always take the last deconcatenation sum to be a 3-deconcatenation, namely the first 3 letters of $P$ must share a common vertex. An example is
\ie
&k_{1234}^2A_{1234\mu}^{\rm contact}=(k_{123\mu}-k_{4\mu})(A^{\rm contact}_{123}\cdot A_{4})\\
&-A^{\rm contact}_{123\mu}(k_{123}\cdot A_{4})-A^{\rm contact}_{123\mu}(k_{1234}\cdot A_{4})\\
&+A_{4\mu}(k_{4}\cdot A^{\rm contact}_{123})+A_{4\mu}(k_{1234}\cdot A^{\rm contact}_{123})\\
&k_{123}^2A^{\rm contact}_{123\mu}=2A_{2\mu}(A_{1}\cdot A_{3})-A_{1\mu}(A_{2}\cdot A_{3})\\
&-A_{3\mu}(A_{1}\cdot A_{2})
\fe
It is instructive to see why the contact component is required at all. If we use only the comb component $A^{\rm comb}_{l12\cdots m}$ in \eqref{bk}, the resulting object always contains a propagator between the leg $l$ and the leg $m$: by the definition of the comb, the last two letters of the word are attached sequentially, so that $l$ and $m$ are separated by at least one internal line. But the one-loop diagram in which the legs $l$ and $m$ meet at a common quartic vertex has no such propagator, and it must be included if the kernel is to be invariant under cyclic permutations of the legs $1,2,\ldots,m$. The contact component is precisely the object that generates this missing class of diagrams: it is defined like the comb component, except that the \textit{last} deconcatenation is forced to be a three-deconcatenation, so that the first three letters of the word share a common quartic vertex. Once the correct one-loop kernel has been constructed, no contact component is needed at higher loops: the subsequent recursion steps generate the higher-loop diagrams from lower-loop kernels whose cyclic invariance is already built in, so the higher-loop kernels are automatically cyclic-invariant.

Also note that after adding the suitable symmetry factors of each term, we will get the 1-loop kernel $I^{\rm kernel}_{1,m}$. In most cases, the 1-loop kernel is easy to obtain:
\ie\label{eq:symm}
&I^{\rm kernel}_{1,m}=\frac{1}{2}I^{\rm bare\ kernel}_{1,m},\ m=2\\
&I^{\rm kernel}_{1,m}=I^{\rm bare\ kernel}_{1,m},\ m\geq5
\fe
while in the cases $m=3$ and $m=4$, only some of the terms should be multiplied by a factor of $1/2$, depending on the types of vertices that appear in the corresponding diagrams. The origin of these factors is discussed in appendix \ref{app:graph}, and in the bi-adjoint scalar theory the same factors can be obtained in an elegant algebraic way from the graded inverse variables of \cite{Tao:2025pnt}. The explicit one-loop kernels for $m=3,4$ are provided as ancillary material, and the two-way kernel, including its ghost part, is displayed explicitly in section \ref{sec5}. We have compared all of them with an independent Feynman-rule computation using FeynCalc \cite{Hahn:2000kx}, finding complete agreement.

\subsection{Loop kernels and the $\ell$-loop planar integrands}
We can obtain higher-loop kernels from the 1-loop kernels, following some recursive steps proposed in \cite{Tao:2025fch}. The main idea of this recursion is to sew comb components with the lower-loop kernels. The recursion for the bi-adjoint scalar theory is reviewed in \cite{Tao:2025fch}, and the YM version, including the ghosts, is given in section \ref{sec4}. After replacing each leg of an $m$-way $\ell$-loop kernel with some BG currents, we will get part of the $\ell$-loop integrands. Such terms actually correspond to irreducible loop diagrams. For the reducible part, we need to consider the loop generalization of the BG currents, namely the loop currents. Formally, the final result for an $\ell$-loop planar integrand is
\ie\label{n-loop}
&I^{\rm \ell-loop}(\pmb{\alpha})=\sum_{k=2}^{n}\sum_{\text{ $k$-division of $\pmb{\alpha}$}}I^{\rm kernel}_{\ell,k}\bigg|_{k_i\to k_{\pmb{\alpha}_i}\atop A_{i}\to \ma{A}_{\pmb{\alpha}_i}}\\
&+\frac{1}{\ell}(\sum_{m=1}^{\ell-1}\sum_{k=2}^{n}\sum_{\ell_i\atop\sum_{i=1}^k \ell_i=\ell-m}\sum_{\text{ $k$-division of $\pmb{\alpha}$}}mI^{\rm kernel}_{m,k}\bigg|_{k_i\to k_{\pmb{\alpha}_i}\atop \phi_{i|i}\to \ma{A}^{(\ell_i)}_{\pmb{\alpha}_i}})
\fe
where the division of $\pmb{\alpha}$, namely $\{\pmb{\alpha}_i\}$, is discussed in the context of cyclic permutation, and $\ma{A}_{\pmb{\alpha}_i\mu}$ is the gluon Berends-Giele current rather than a multi-particle current.

The two lines of \eqref{n-loop} have a clear physical meaning. The first line produces the \textit{irreducible} part of the integrand: an $\ell$-loop kernel with all $\ell$ loops connected, whose legs are replaced by Berends-Giele currents. The second line produces the \textit{reducible} part, in which the diagram factorizes into several components with fewer loops when a single propagator is cut; these components are connected through the loop kernels $I^{\rm kernel}_{m,k}$ of the first sum, while the loop currents $\ma{A}^{(\ell_i)}$ represent the off-shell components that contain the remaining loops. The factor $m/\ell$ removes the overcounting between the different $m$-loop components of a reducible diagram: a reducible diagram containing $n_m$ components with $m$ loops each satisfies $\sum_m m\,n_m=\ell$, and it is generated $n_m$ times from the term with $m$-loop kernels; multiplying by $m/\ell$ and summing over $m$ gives $\sum_m n_m\, m/\ell=1$, i.e.\ each diagram is counted exactly once. The symmetry factors of the diagrams affect only the recursion that constructs the loop kernels and are not modified by the substitutions of \eqref{n-loop}; we discuss them in appendix \ref{app:graph}.

\section{Matrix formalism of the pure gluon sector}\label{sec3}
\subsection{Transfer-matrix representation of the comb}
In this section we focus on the pure-gluon sector and set all ghost currents to zero. The key observation is that the comb component defined in section \ref{sec2} satisfies a very special recursion: because a new letter is always attached at the moving end of the comb, the current of the word $P(m)$ is determined by the currents of the two preceding words $P(m-1)$ and $P(m-2)$ through a linear relation with \textit{known} coefficient matrices, which depend on the word $P(m)$ but not on any of the earlier words $P(1),\ldots,P(m-1)$. To see this, we simply restrict the gluon recursion \eqref{gluoncurrent} to the comb: keeping only the splittings that define the comb component, the $m$-th step reads
\ie\label{eq:comb-bc}
A^{\rm comb}_{P(m)}\cdot\epsilon=&\epsilon_{\mu}B_{\mu\nu}(P(m))A^{\rm comb}_{P(m-1)\nu}\\
&+\epsilon_{\mu}C_{\mu\nu}(P(m))A^{\rm comb}_{P(m-2)\nu}
\fe
where $P(i)$ is the word with $P_i$ being the last letter of it, i.e. $P(i)=P_1P_2\cdots P_i$, and
\ie
&B_{\mu\nu}(P(m))\\
&=\frac{(k_{P(m-1)\mu}-k_{P_m\mu}) A_{P_m\nu}-\eta_{\mu\nu}(k_{P(m-1)}\cdot A_{P_m})}{k_{P(m)}^2}\\
&+\frac{-\eta_{\mu\nu}(k_{P(m)}\cdot A_{P_m})+A_{P_m\mu}k_{P_m\nu}+A_{P_m\mu}k_{P\nu}}{k_{P(m)}^2}\\
&C_{\mu\nu}(P(m))\\
&=\frac{2A_{P_{m-1}\mu} A_{P_m\nu}-\eta_{\mu\nu}(A_{P_{m-1}}\cdot A_{P_m})-A_{P_m\mu}A_{P_{m-1}\nu}}{k_{P(m)}^2}\,.
\fe

The recursion \eqref{eq:comb-bc} has the same shape as the Fibonacci recurrence $f_m=f_{m-1}+f_{m-2}$, except that the coefficients are matrices acting on Lorentz vectors. Any linear recurrence of this two-step form can be rewritten as a one-step recursion on a doubled state: if we collect the current of the word $P(m)$ and that of the preceding word $P(m-1)$ into a two-component object, the step $m-1\to m$ acts by a single $2\times2$ block matrix, the \textit{transfer matrix}
\ie\label{eq:transfer}
D(P(m))\equiv\begin{pmatrix}
B(P(m)) & C(P(m))\\
1 & 0
\end{pmatrix},
\fe
where the entries $1$ and $0$ denote the identity and the zero operators on Lorentz vectors. Iterating from the seed $P(1)$ then gives the following closed formula:
\begin{widetext}
\ie\label{pgcomb}
\bigg(\begin{array}{c}A^{\rm comb}_{P(m)}\cdot\epsilon\\A^{\rm comb}_{P(m-1)}\cdot\epsilon\end{array}\bigg)=
\epsilon\cdot\begin{pmatrix}
B(P(m)) & C(P(m))\\
1 & 0
\end{pmatrix}\cdot\ldots\cdot\begin{pmatrix}
B(P(2)) & C(P(2))\\
1 & 0
\end{pmatrix}\cdot\bigg(\begin{array}{c}\epsilon_{P_1}\\ 0\end{array}\bigg)
\fe
\end{widetext}
where the label \eqref{eq:comb-bc} refers to the recursion displayed at the beginning of this subsection. Because each factor $D(P(i))$ is built only from the momenta and single-particle currents associated with the letters $P_{i-1},P_i$ (through $B$ and $C$) together with the accumulated momentum $k_{P(i)}$, the formula makes the local structure of the construction manifest at the level of the word: the current of a long word is assembled one vertex at a time.

Before proving \eqref{pgcomb}, we fix the index conventions precisely, since they are the main source of potential confusion in a block-matrix formula whose entries are tensors. The dot ``$\cdot$'' between two neighboring factors of \eqref{pgcomb} means both matrix multiplication and contraction of the Lorentz indices of the two adjacent tensor factors. The seed column $(\epsilon_{P_1}\,;\,0)^T$ is a column of Lorentz vectors whose upper entry is the single-particle current $A_{P_1\mu}=\epsilon_{P_1\mu}$; the lower entry is the zero vector. The chain of transfer matrices acts from the right on this seed, producing a new column of Lorentz vectors whose upper entry is $A^{\rm comb}_{P(m)\mu}$; the leading polarization $\epsilon$ is contracted with the free Lorentz index of this upper entry only, which is the meaning of writing $\epsilon\cdot$ in front of the product. In particular, the lower entry of the final column is $A^{\rm comb}_{P(m-1)\mu}$, as indicated on the left-hand side. Equivalently, the identity operator in the lower-left entry of $D(P(i))$ is $\delta_\mu{}^\nu$ in Lorentz space.

To construct the one-loop kernel, we need to set the first letter of the comb to be the loop leg $l$. We denote the word $lP_1\cdots P_i$ by $P(l,i)$, so that $P(l,0)=l$ and $P(l,m)=lP_1\cdots P_m$; the comb components of these words follow from \eqref{pgcomb} with the replacement $P(m)\to P(l,m)$ and the seed $(\epsilon_l;0)^T$.

\subsection{The contact component and the one-loop kernel}
To make the 1-loop kernel cyclic invariant, we also need to add the contact component $A^{\rm contact}_{lP_mP_1\cdots P_{m-1}}$. The transfer matrices that appear in its recursion are different from the ones of the comb, because the word of the contact component starts with the pair $lP_m$ rather than with the single letter $l$. However, the two sets of matrices are related by a simple momentum shift, and this is the mechanism that ultimately puts the contact component into the same transfer-matrix language. To see this, note that the matrices $B$ and $C$ of a given word depend on the word only through the momentum accumulated before its last letter, the total momentum of the word, and the single-particle currents of the last two letters. For the word $lP_mP_1\cdots P_i$ the momentum accumulated before the last letter $P_i$ is $k_l+k_{P_m}+k_{P_1\cdots P_i}$ (with the obvious adjustment for the quartic part, which involves the last two letters), while for the word $P(l,i)=lP_1\cdots P_i$ it is $k_l+k_{P_1\cdots P_i}$. Hence, under the momentum shift $l\to l-k_{P_m}$ the two agree, and we find that
\ie
B_{\mu\nu}(lP_mP_1\cdots P_i)&\to B_{\mu\nu}(P(l,i))\\
C_{\mu\nu}(lP_mP_1\cdots P_i)&\to C_{\mu\nu}(P(l,i)).
\fe
Physically, the shift moves the leg $P_m$ from the beginning of the chain to the quartic vertex at which it meets the leg $P_1$; the momentum of $P_m$ no longer flows through any of the transfer matrices of the chain. The contact component can therefore be written in exactly the same form as the comb component, with two differences: the chain now contains only the matrices of the words $P(l,2),\ldots,P(l,m-1)$, and the seed is no longer a single-particle current but the matrix $M(P(m))$, which encodes the quartic vertex containing the legs $l$, $P_m$ and $P_1$. Concretely,
\begin{widetext}
\ie
\bigg(\begin{array}{c}A^{\rm contact}_{lP_mP_1\cdots P_{m-1}}\cdot\epsilon\\A^{\rm contact}_{lP_mP_1\cdots P_{m-2}}\cdot\epsilon\end{array}\bigg)
=&\,\epsilon\cdot\begin{pmatrix}
B(P(l,m-1)) & C(P(l,m-1))\\
1 & 0
\end{pmatrix}\cdot\ \cdots\\
&\quad\cdot\begin{pmatrix}
B(P(l,2)) & C(P(l,2))\\
1 & 0
\end{pmatrix}\cdot\bigg(\begin{array}{c}M(P(m))\\ 0\end{array}\bigg)\cdot\epsilon_l
\fe
\end{widetext}
with 
\ie
M_{\mu\nu}(P(m))=&\,\frac{2A_{P_m\mu}A_{P_1\nu}-\eta_{\mu\nu}(A_{P_m}\cdot A_{P_1})}{(l+k_{P_1})^2}\\
&-\frac{A_{P_1\mu}A_{P_m\nu}}{(l+k_{P_1})^2}\,.
\fe

The denominator $(l+k_{P_1})^2=k_{P(l,1)}^2$ is the first propagator of the contact chain, i.e.\ the propagator that separates the quartic vertex from the rest of the tree.

We are now in a position to assemble the one-loop kernel. Recall from \eqref{bk} that the bare kernel is obtained from $k_{P(l,m)}^2(A^{\rm comb}_{P(l,m)}+A^{\rm contact}_{lP_mP_1\cdots P_{m-1}})\cdot\epsilon$ by the sewing replacement $\epsilon_{l\mu}\epsilon_\nu\to\eta_{\mu\nu}/l_1^2$. In the transfer-matrix language the recursion \eqref{eq:comb-bc} is already solved: the comb current of the word $P(l,m)$ is the upper component of the state \eqref{eq:state-sol} with seed $(\epsilon_l;0)^T$, and the corresponding contact current is the upper component of the chain with seed $(M(P(m));0)^T$ discussed above. What remains is to implement the sewing: the loop leg $l$ is turned into an internal line by the replacement $\epsilon_{l\mu}\epsilon_\nu\to\eta_{\mu\nu}/l_1^2$, which removes the two remaining uncontracted Lorentz indices of the two chains and explains the overall metric contraction $\eta^{\mu\nu}$ in the formula below. The last recursion step, which builds the two-letter word $P(l,1)=lP_1$ out of the seed $A_l=\epsilon_l$, is a special case of the same structure and is conveniently packaged into the two-component column $(B(P(l,1))\,;\,1)^T$ contracted with $\epsilon_l$; the loop propagators that survive the amputation by $k_{P(l,m)}^2$ are contained in the denominators of the matrices $B,C,M$. Iterating the chains and applying the sewing therefore gives the following closed expression for the YM one-loop bare kernel:
\begin{widetext}
\ie\label{1loopkernel}
\bigg(\begin{array}{c}I^{\rm bare\ kernel}_{1,m}\\ \#\end{array}\bigg)
=&\ \eta^{\mu\nu}\biggl[\begin{pmatrix}B(P(l,m)) & C(P(l,m))\\1 & 0\end{pmatrix}\\
&\qquad\cdot\ \cdots\ \cdot\begin{pmatrix}B(P(l,2)) & C(P(l,2))\\1 & 0\end{pmatrix}
\cdot\begin{pmatrix}B(P(l,1))\\1\end{pmatrix}\biggr]_{\mu\nu}\\
&+\eta^{\mu\nu}\biggl[\begin{pmatrix}B(P(l,m-1)) & C(P(l,m-1))\\1 & 0\end{pmatrix}\\
&\qquad\cdot\ \cdots\ \cdot\begin{pmatrix}B(P(l,2)) & C(P(l,2))\\1 & 0\end{pmatrix}\cdot\bigg(\begin{array}{c}M(P(m))\\ 0\end{array}\bigg)\biggr]_{\mu\nu}
\fe
\end{widetext}
The notation $\#$ means some messy terms we do not care about. After the metric contraction of the sewing this lower component does not contribute to the $m$-way kernel and is therefore dropped. The $1$-loop kernel $I^{\rm kernel}_{1,m}$ is obtained from the bare kernel by multiplying each term with its symmetry factor, as discussed around \eqref{eq:symm}; to obtain the one-loop integrands, one replaces each single-particle current $A_{P_i}$ by a Berends-Giele current, exactly as described in section \ref{sec2}.

\subsection{Some remarks}
The advantage of formula \eqref{1loopkernel} is that it separates the contributions of different vertices and different legs. Each transfer matrix in \eqref{1loopkernel} represents one vertex of the loop, and a given leg $P_i$ with $i\neq m$ appears only in the two adjacent matrices of each term; the leg $P_m$ appears only in $M(P(m))$, $B(P(l,m))$, and $C(P(l,m))$. This local structure has an immediate computational consequence. Consider a term of a higher-loop kernel that is obtained by sewing a comb component into a one-loop kernel built from \eqref{1loopkernel}, i.e.\ by replacing one single-particle current $A_{P_i}$ (and the momentum $k_{P_i}$) with the current of a longer word. In the chain product, this replacement modifies at most two adjacent transfer matrices. Re-evaluating the term therefore requires only the local substitution of those matrices and a re-multiplication of the chain, which involves a number of matrix multiplications proportional to the length $m$ of the chain, rather than a re-evaluation of all Feynman diagrams with the new subcurrent. In contrast, in a conventional Feynman-rule evaluation the same substitution forces one to re-expand the whole diagram in terms of the new vertices and to redo the full sum over graphs. This is the sense in which the matrix formalism improves the computational organization of the recursion: the sewing is a local operation on the chain, both at the level of the formulas and at the level of an implementation, and the same property persists at higher loops because every higher-loop kernel is built from one-loop kernels of this form.

Moreover, the comb component \eqref{pgcomb} is also written in the matrix formalism. In the bi-adjoint case, the sewing procedure in the loop kernel recursion is rather easy, and it can be regarded as sewing a propagator chain to the lower-loop kernel. However, in the YM case, we should regard the same process as sewing a matrix chain to the lower-loop kernel. For example, if we use the notation
\ie
D_{\mu\nu}^{l_1,i}\equiv\begin{pmatrix}
B(P(l_1,i)) & C(P(l_1,i))\\
1 & 0
\end{pmatrix},
\fe
then sewing the comb components, say $A_{a_212\mu}^{\rm comb}$, with leg $P_2$ of a 4-way 1-loop gluon kernel can be expressed as
\ie
I^{\rm kernel}_{1,4}\supset(D_{\mu\nu}^{l_1,3}D_{\nu\rho}^{l_1,2})\bigg|_{A_{P_2\mu}\to A_{a_212\mu}^{\rm comb}}, k_{P_2}\to l_2+k_1+k_2
\fe
where $D_{\mu\nu}^{l_1,3}D_{\nu\rho}^{l_1,2}$ is the only place where $A_{P_2\mu}$ will appear.

The structure of the transfer matrices is also what makes the one-loop differential-operator relations between the Yang-Mills integrands and their scalar-loop counterparts transparent in the present formalism \cite{Chen:2023bji,Zhou:2021kzv,Zhou:2022djx,Cheung:2017ems}. The relevant observation is simple. In an expression such as \eqref{1loopkernel}, factors of the spacetime dimension $d$ are generated only through the contraction of the overall metric with a metric tensor coming from the interior of the chain, $\eta^{\mu\nu}\eta_{\mu\nu}=d$; all other Lorentz contractions involve momenta or polarization vectors and produce no factor of $d$. A derivative with respect to the dimension, $\partial_d$, therefore acts locally on the chain: it selects, inside every matrix of \eqref{1loopkernel}, the part proportional to the metric,
\ie\label{eq:metric-bc}
&B^{s}_{\mu\nu}(P(l,i))=\frac{-\eta_{\mu\nu}\left(k_{P(l,i-1)}\cdot A_{P_i}\right)-\eta_{\mu\nu}\left(k_{P(l,i)}\cdot A_{P_i}\right)}{k_{P(l,i)}^2},\\
&C^{s}_{\mu\nu}(P(l,i))=-\frac{\eta_{\mu\nu}\left(A_{P_{i-1}}\cdot A_{P_i}\right)}{k_{P(l,i)}^2},
\fe
and it turns the contact seed into
\ie\label{eq:metric-m}
M^{s}_{\mu\nu}(P(m))=-\frac{\eta_{\mu\nu}\left(A_{P_m}\cdot A_{P_1}\right)}{(l+k_{P_1})^2}.
\fe
These elements are exactly what we meet in the transfer-matrix representation of the single-trace YMS comb components with all legs gluons except $l$ and $P(m)$; hence this kind of comb component corresponds to a YMS 1-loop kernel with a scalar running in the loop, whose external legs are all gluons.

Finally, we remark on the relation of the transfer-matrix formalism to amplitude relations. Since each leg appears in a fixed, small number of neighboring matrices, any operation on the integrand that acts on a single leg---such as a differential operator that transmutes one theory into another, or a momentum shift that identifies the leg with a different one---has an action that reaches only a small number of adjacent matrices of the chain. This is precisely the property that is needed to transport the tree-level relations among off-shell currents \cite{Chen:2023bji,Tao:2022nqc,Tao:2024vcz,Wu:2021exa,Tao:2023yxy} through the sewing procedure to the loop level in a controlled way, and it is the reason we expect the formalism to be useful for deriving higher-loop amplitude relations. We leave a systematic study of such relations to future work.

\section{The ghost sector and the complete recursion}\label{sec4}
\subsection{Ghost currents and their comb components}
In this section we extend the recursion of sections \ref{sec2} and \ref{sec3} to the complete YM theory by including the Faddeev-Popov ghosts. The ghosts are required for the consistency of the off-shell construction: in the Feynman gauge the gluon field has unphysical polarizations, and the ghosts cancel their contributions inside every loop. In the recursive language of the multi-particle currents this shows up in two places: the ghost currents themselves, and the ghost terms inside the gluon recursion \eqref{gluoncurrent}, which are generated by the gluon-ghost-antighost vertex.

The ghost multi-particle currents $b_P$ and $c_P$ are defined by the same perturbiner expansion as the gluon currents, and they satisfy the algebraic recursions
\ie\label{ghostcurrent}
k_{P}^2 b_{P}&=k_{Y}\cdot A_{X} b_{Y}-k_{X}\cdot A_{Y} b_{X}\\
-k_{P}^2 c_{P}&=-k_{P}\cdot A_{X}c_{Y}+k_{P}\cdot A_{Y} c_{X}
\fe
with the sums over the two-deconcatenations $P=XY$ left implicit, as in \eqref{gluoncurrent}. These recursions are obtained from the ghost equations of motion $\partial^2 c\sim\partial^\mu[A_\mu,c]$, $\partial^2 b\sim\partial^\mu[A_\mu,b]$ in exactly the same way as \eqref{gluoncurrent}; because the ghosts only interact through the derivative cubic coupling with the gluons, no quartic ghost vertex exists and the recursion contains only two-deconcatenations. The ghost terms in the gluon recursion \eqref{gluoncurrent} must be kept simultaneously: they are the last line of that equation. Note that from the Lagrangian we have 
\ie\label{eq:ghost-props}
\langle A\,A\rangle&\sim \frac{1}{k^2}\\
\langle c\,b\rangle&\sim \frac{1}{k^2}\\
\langle b\,c\rangle&\sim -\frac{1}{k^2}
\fe
where the correlation functions are read with the fields in the indicated order, left to right. The sign of $\langle b\,c\rangle$ is the content of the statement that the ghost propagator is orientation-dependent: a $c$ leg sewn to a $b$ leg and a $b$ leg sewn to a $c$ leg carry opposite signs. In the recursion \eqref{ghostcurrent} this orientation dependence is encoded by writing the equation for $c$ with a minus sign on the left-hand side, i.e.\ $k_P^2c_P=k_P\cdot A_X c_Y-k_P\cdot A_Y c_X$; equivalently, the $c$ multi-particle current carries an extra overall minus relative to the naive extension of the gluon recursion, which is precisely the sign needed to reproduce $\langle b\,c\rangle=-\langle c\,b\rangle$ after sewing.

The comb components of the ghost currents are defined exactly as in the gluon case: we keep only the term in which the last letter is separated from the rest, i.e.\ $X=P(m-1)$, $Y=P_m$ in \eqref{ghostcurrent}. Restricting the recursion accordingly, we obtain the explicit comb recursions
\ie\label{eq:ghostcomb}
&k_{P(m)}^2 b^{\rm comb}_{P(m)}=k_{P_m}\cdot A^{\rm comb}_{P(m-1)}\,b_{P_m}\\
&\qquad\qquad-k_{P(m-1)}\cdot A_{P_m}\,b^{\rm comb}_{P(m-1)},\\
&k_{P(m)}^2 c^{\rm comb}_{P(m)}=k_{P(m)}\cdot A^{\rm comb}_{P(m-1)}\,c_{P_m}\\
&\qquad\qquad-k_{P(m)}\cdot A_{P_m}\,c^{\rm comb}_{P(m-1)},
\fe
whose structure is the same as that of \eqref{gluoncurrent} restricted to the two-deconcatenation part. The only difference between $b$ and $c$ is the momentum contracted with the gluon current in the first term: $k_{P_m}$ (the momentum of the newly attached leg) for $b$, and $k_{P(m)}$ (the total momentum of the word) for $c$; this difference is forced by the equations of motion and is ultimately responsible for the relative signs of the two ghost orientations in the sewing rules below. A concrete illustration is provided by the three-letter examples
\ie
&k_{123}^2 b_{123}^{\rm comb}=k_{3}\cdot A_{12}^{\rm comb} b_{3}-k_{12}\cdot A_{3} b_{12}\\
&k_{123}^2 c_{123}^{\rm comb}=k_{123}\cdot A_{12}^{\rm comb} c_{3}-k_{123}\cdot A_{3} c_{12}
\fe
which follow directly from \eqref{eq:ghostcomb}. Note that the single-particle states of the ghost currents, $b_i$ and $c_i$, are anticommuting variables; they vanish when the corresponding leg is put on shell. The consequences of this anticommutation for the ordering of the fields inside the recursion are discussed in the next subsection.

\subsection{sewing rules}
Because the ghost fields are Grassmann-odd, every formula in this paper that involves ghosts must be read as an identity in the graded (super) algebra generated by the fields, and sign conventions must be fixed once and for all. We use the following rules, which are the minimal set needed to make the recursion unambiguous:
\begin{enumerate}
\item The gluon fields $A_{P\mu}$ are Grassmann-even; the ghost fields $b_P$, $c_P$ and the single-particle symbols $b_i$, $c_i$ are Grassmann-odd.
\item The ghost charge is conserved by the interaction: we assign charge $+1$ to $c$ and $-1$ to $b$, and every vertex of the Lagrangian, including the gluon-ghost vertex of the last line of \eqref{gluoncurrent}, carries zero total ghost charge. Consequently, every vertex of the recursion carries zero net ghost charge, and a closed ghost line always contains equal numbers of $b$ and $c$ insertions; no physical gluonic object ever carries an open ghost line, a property that is checked explicitly at the end of each construction step.
\item Within a multi-particle current, the single-particle ghost symbols are ordered according to the word: reading a word $P=p_1\cdots p_m$ from left to right, the corresponding fields appear in the order $p_1,\ldots,p_m$. Whenever two ghost symbols appear in an order that is not the word order, the equation has been obtained by moving them past each other and the associated sign has already been included. In particular, the products $b_{Y}c_{X}$ and $c_{X}b_{Y}$ in equations such as the last line of \eqref{gluoncurrent} differ by a sign, and only one of the two orderings is written.
\item The sewing of two legs is performed by bringing the two contracted fields to adjacent positions in the monomial (using rule 1, keeping the signs), and then applying the replacements
\ie\label{eq:sewing}
\epsilon_{l\mu}\epsilon_\nu\to\frac{\eta_{\mu\nu}}{l_1^2},\qquad
c\,b_l\to\frac{1}{l_1^2},\qquad
b\,c_l\to-\frac{1}{l_1^2},
\fe
where $b_l$, $c_l$ denote the single-particle ghost symbols of the leg $l$ that is turned into an internal line, and the fields on the left of these symbols are the single-particle ghosts of the legs to which the leg $l$ is sewn. Rule 4 is the translation of the propagator signs of \eqref{eq:ghost-props} into the recursion: it is the only place in the whole construction where an explicit minus sign for the orientation of a ghost line is introduced.
\end{enumerate}
With these conventions, no further sign bookkeeping is needed: the recursion and the sewing are performed in the graded algebra, and all signs are fixed by the anticommutation rule and by the replacements \eqref{eq:sewing}. We will see a nontrivial example of this bookkeeping in section \ref{sec5}, where the two-loop kernel with a ghost loop is assembled; there the cyclic ordering of the two orientations $c\,b$ and $b\,c$ produces, through \eqref{eq:sewing}, the overall minus sign of the fermion loop.

\subsection{Ghost contributions to the one-loop kernels}
We now construct the one-loop kernels including the ghosts. To construct the $m$-way 1-loop bare kernel, we have three contributions:
\begin{enumerate}
\item From $k^2_{12\cdots m}A^{\rm comb}_{l12\cdots m}\cdot\epsilon_n+k^2_{12\cdots m}A^{\rm contact}_{lm12\cdots m-1}\cdot\epsilon$ with $l$ a gluon leg, using $\epsilon_{l\mu}\epsilon_{\nu}\to \eta_{\mu\nu}/l_1^2$ as before.

\item From $k^2_{12\cdots m}c\,b^{\rm comb}_{lP}$ with $l$ a $b$ ghost leg, using $c\,b_l\to 1/l_1^2$.

\item From $-k^2_{12\cdots m}b\,c^{\rm comb}_{lP}$ with $l$ a $c$ ghost leg, using $b\,c_l\to -1/l_1^2$.
\end{enumerate}
In the ghost sewing procedure, the single-particle ghost of the loop leg, $b_l$ or $c_l$, must be in the first position of the corresponding comb current; for example $b^{\rm comb}_{lP}$ is written with the $b_l$ factor leftmost, $b_{lP}=b_l\,\ma{B}_P$, where $\ma{B}_P$ is a Grassmann-even function of the gluon currents and of the remaining ghost insertions. This convention is compatible with rule 3 above because the loop leg is, by definition, the first letter of the word.

There is one further subtlety, which is the ghost analog of the symmetry factors of the gluon kernels: for a two-way kernel part that contains both ghost and gluon propagators, cases 2 and 3 above, together with case 1, generate the \textit{same} loop diagram, and the three contributions must therefore be weighted by a factor of $1/2$ to avoid overcounting, exactly as in the bi-adjoint scalar case \cite{Tao:2025pnt}. Together with the symmetry factor of the two-way kernel itself, this produces the $1/2$ factors of the explicit formulas below. We have computed the complete one-loop kernels including the ghosts for $m=2,3,4$; the $m=2$ kernel is displayed in section \ref{sec5}, and the $m=3,4$ kernels are provided as ancillary material together with the pure-gluon kernels. As a nontrivial validation of the ghost recursion and of the sign rules of the previous subsection, we have compared the complete kernels (gluon and ghost parts together) against an independent Feynman-rule computation performed with FeynCalc \cite{Hahn:2000kx}, finding complete agreement for all three values of $m$ up to some shifts of the momenta; the gluon-only parts of the same comparison were already reported in \cite{Tao:2025fch}.

\subsection{Loop-kernel recursion with ghosts}
When we go to higher loops, we need to consider all the sewing cases. Compared with the pure gluon case, there are some extra cases: we also need to consider $b^{\rm comb}_{\cdots}$ and $c^{\rm comb}_{\cdots}$. For each leg of the lower-loop kernel, we sew the sum of all comb components $A^{\rm comb}_{\cdots\mu}+b^{\rm comb}_{\cdots}+c^{\rm comb}_{\cdots}$ rather than only $A^{\rm comb}_{\cdots\mu}$. This is the only modification of the recursion rules of section \ref{sec2}; the topological bookkeeping, i.e.\ the ordered sets, the $k$-divisions, and the graph factors, is unchanged, since it only depends on the loop topology and not on the particle content of the lines.

We now conclude the new loop-kernel recursion steps for the complete YM theory:
\begin{enumerate}
\item Consider ordered cyclic set $(b_\ell ,a_\ell,1,2, \cdots ,m_\ell-1,m_\ell)$ and the cyclic permutation of $(12\cdots m_\ell)$, i.e. there are $m_\ell$ sets in total. Note that due to the cyclic properties, $(1,2,\cdots,m_\ell-1,m_\ell,a_\ell,b_\ell)$ is equivalent to $(a_\ell,b_\ell,1,2\cdots ,m_\ell-1,m_\ell)$. 
\item Then, for $k$-divisions, which means we divide a set into $k$ parts, of each set, we set: i) $b_{\ell}$ itself to be one part of the $k$-division, ii) For a part not involving $a_{\ell}$, there can only be 1 element in the part, like $(b_{\ell}|a_{\ell},1|2|3)$. There is only 1 case for a given $k$ and a given set. 
\item Consider a $k$-way bare $(\ell-1)$-loop kernel, replace $A_{i\mu}$, $b_i$, $c_i$ of this kernel with the comb component $A^{\rm comb}_{\cdots\mu}$, $b^{\rm comb}_{\cdots}$, $c^{\rm comb}_{\cdots}$ of each part of a $k$-division respectively. Then do the replacement: $A_{a_\ell\mu}A_{b_\ell\nu}\to\eta_{\mu\nu}/l_\ell^2$, $c_{b_\ell}b_{a_\ell}\to1/l_\ell^2$, and $b_{b_\ell}c_{a_\ell}\to-1/l_\ell^2$ together with $k_{a_\ell}=-k_{b_\ell}=l_\ell$. This replacement is equivalent to turning legs $a_\ell$ and $b_\ell$ into an internal line. Then add the corresponding graph factor $g$ to each case. Since we have demonstrated the construction of the Feynman diagrams, one can easily draw diagrams according to the steps above and find the corresponding graph factors.
\item Sum over all possible $k$-division and all $k\geq 2$. Finally we will obtain a $m_\ell$-way $\ell$-loop kernel from $(\ell-1)$-loop kernel $I^{\rm kernel}_{\ell,m_\ell}$.
\end{enumerate}
After getting loop kernels, we can obtain the planar integrands just as \eqref{n-loop}. Note that all loop kernels appearing in the integrand construction formula must have all legs gluon to make sure that no ghost legs are in the final integrands.

Two comments are in order about the role of the graph factor $g$ in the ghost sector. First, the factor $g=S\times(\ell-r)^{-1}$ of appendix \ref{app:graph} is a purely topological quantity: it is determined by the loop structure of the diagram and is independent of whether a given line is a gluon or a ghost. Second, every closed ghost loop carries an extra factor of $-1$ (the fermion-loop sign). In our recursion this sign should be put in by hand. We will show examples later.

\section{Recursion strategy for 2-loop planar integrands}\label{sec5}
In this section we apply the recursion of the previous sections to a concrete case, the two-loop planar integrands, as an example. The two-loop case is rich enough to exhibit all the ingredients of the general construction---the $k$-divisions, the graph factors, and the two ghost orientations---while remaining simple enough that every step can be spelled out. We first consider the two-point two-loop integrand to build intuition, and then conclude a general recursion strategy. Following \eqref{n-loop}, the two-point two-loop integrand is
\ie\label{2-loop}
&I^{\rm 2-loop}(12)=I^{\rm kernel}_{2,2}\bigg|_{k_i\to k_{P_i}\atop A_{i}\to \ma{A}_{P_i}}+\frac{1}{2}\sum_{\ell_i\atop \sum_{i}\ell_i=1}I^{\rm kernel}_{1,2}\bigg|_{k_i\to k_{P_i}\atop A_i\to \ma{A}^{(\ell_i)}_{P_i}}
\fe
with $P=12$. The first term is the irreducible part, built from the two-way two-loop kernel $I^{\rm kernel}_{2,2}$ with the single-particle currents replaced by Berends-Giele currents; the second term is the reducible part, built from the two-way one-loop kernel $I^{\rm kernel}_{1,2}$ in which one leg carries a one-loop current $A^{(1)}$. We now construct each of these objects.

\subsection{Two-way one-loop kernels and one-loop currents}
Let us first consider the two-way one-loop kernel. Its pure-gluon part was given in \cite{Tao:2025fch}; we denote it by $I_{1,2}^{\rm gluon}$. The ghost contributions are
\ie\label{eq:oneloop-ghost}
I^{\rm ghost}_{1,2}=&-\frac{1}{2l^2k_{l1}^2}k_{l}\cdot A_{2} k_{l1}\cdot A_{1}-\frac{1}{2l^2k_{l1}^2}k_{l1}\cdot A_{2}k_{l}\cdot A_{1}\\
&-\frac{1}{l^2k_{l1}^2}k_{l}\cdot k_{1}b_{1} c_{2}-\frac{1}{l^2k_{l1}^2}k_{2}\cdot k_{l}c_{1} b_{2}\\
&-\frac{1}{2l^2k_{l1}^2}k_{2}\cdot k_{1}c_{1} b_{2}-\frac{1}{2l^2k_{l1}^2}k_{1}\cdot k_{1}b_{1} c_{2}
\fe
where we have used the two-point kinematics $k_{12}=0$. This expression contains all three types of contributions listed in section \ref{sec4}: the first two terms come from the ghost part of the gluon recursion (case 1 of that section), the middle two terms from the pure ghost loops with a $b$ or $c$ loop leg (cases 2 and 3), and the last two terms from the mixed diagrams that are counted twice by cases 1--3 and hence carry the factor $1/2$; an overall minus sign has been included for the terms that close a fermion (ghost) loop, according to the sign rules of section \ref{sec4}. The total two-way one-loop kernel is
\ie
I^{\rm kernel}_{1,2}=I_{1,2}^{\rm gluon}+I^{\rm ghost}_{1,2}
\fe
and the one-point one-loop gluon current, needed for the reducible part of \eqref{2-loop}, is obtained from it by putting leg 1 on shell:
\ie\label{eq:oneloop-current}
A^{(1)}_{1}\cdot A_2=&I_{1,2}^{\rm gluon}\bigg|_{\text{leg 1 on-shell}\atop k_2\to-k_1}\\
&-\frac{1}{2l^2k_{l1}^2}k_{l}\cdot A_{2} k_{l}\cdot A_{1}-\frac{1}{2l^2k_{l1}^2}k_{l1}\cdot A_{2}k_{l}\cdot A_{1}
\fe
Here $A_2$ does not satisfy any transversality condition; note that only the pure-gluon part of $I^{\rm kernel}_{1,2}$ survives in the first term after leg 1 is put on shell, because the ghost single-particle states vanish on shell. Hence the second term of \eqref{2-loop} is fully determined at this point.

\subsection{Two-way two-loop kernel with ghosts}
As for the first term of \eqref{2-loop}, we need to construct the two-way two-loop kernel $I^{\rm kernel}_{2,2}$ of the complete YM theory. According to the recursion of section \ref{sec4}, this kernel involves the two-, three- and four-way one-loop kernels. It is important to understand which of them are actually needed, since this determines the computational cost of the recursion. We now spell out the relevant bookkeeping. For the two-way kernel, the ordered cyclic sets of the recursion are $(b_2,a_2,1,2)$ and $(b_2,a_2,2,1)$ together with the cyclic permutations of $(12)$; since $(12)$ has only one nontrivial cyclic image, namely $(21)$, this gives the two sets just written. For each set, the constraints of the recursion leave exactly one $k$-division for each $k=2,3,4$:
\begin{align*}
k=2:&\quad (b_2\,|\,a_2,1,2),\quad (b_2\,|\,a_2,2,1),\\
k=3:&\quad (b_2\,|\,a_2,1\,|\,2),\quad (b_2\,|\,a_2,2\,|\,1),\\
k=4:&\quad (b_2\,|\,a_2\,|\,1\,|\,2),\quad (b_2\,|\,a_2\,|\,2\,|\,1),
\end{align*}
so that the $k$-way one-loop kernels with $k=2,3,4$ enter, as expected. Sewing the comb components of the parts into the corresponding $k$-way one-loop kernels and applying the replacements of step 3 of the recursion (together with the graph factors of appendix \ref{app:graph}) produces, after summing over the two cyclic sets, the two-way two-loop kernel. In most cases the terms produced from the four-way and from the two-way one-loop kernels coincide after the sewing and the momentum relabeling, the only exception being the terms in which only four-point vertices appear. Taking into account the graph factors, the most economical way to compute the two-way two-loop kernel is therefore to sew comb components into the combination
\ie
I^{\rm ghost}_{1,2}+\frac{1}{2}I^{\rm ghost}_{1,3}+\frac{1}{2}\sum_{i=2}^{4}I^{\rm gluon}_{1,i}\xrightarrow{\text{sew comb components}}I^{\rm kernel}_{2,2}
\fe
where $I^{\rm gluon}_{1,i}$ and $I^{\rm ghost}_{1,i}$ denote, respectively, the pure-gluon part and the ghost part of the $i$-way one-loop kernel, $I^{\rm ghost}_{1,i}=I^{\rm kernel}_{1,i}-I^{\rm gluon}_{1,i}$. We emphasize that this reduction is a property of the two-loop topology: it is the explicit form of the statement, made in section \ref{sec2} and appendix \ref{app:graph}, that the graph factor accounting can be organized by the implicit symmetries of the construction, so that not all $k$-way kernels have to be processed independently.

As for the loop kernel recursion itself, all steps are the same as in section \ref{sec4}, with one restriction: since we only want two-loop integrands whose legs are gluons, when we sew comb components into one-loop kernels we must choose the first leg of a ghost comb component to be a ghost leg (rather than a gluon leg), and for the gluon comb components we may use the pure-gluon sector of section \ref{sec3}. Under this restriction only two kinds of one-loop kernels contribute: the kernels with all-gluon legs, and the kernels with all but two adjacent ghost legs. The all-gluon case follows directly from the formalism of section \ref{sec3}. For the kernels with two adjacent ghost legs, two orientations are possible, according to the cyclic order of the two ghosts: reading clockwise, the pair can be $(c,b)$ or $(b,c)$. In the former case we replace the $c$ leg by the comb component $c^{\rm comb}(b_2)$ and the $b$ leg by $b^{\rm comb}(a_2,\cdots)$, and then sew the pair with $c_{b_2}b_{a_2}\to1/l_2^2$; in the latter case the roles of $b$ and $c$ are exchanged and the sewing uses $b_{b_2}c_{a_2}\to-1/l_2^2$. After both orientations have been processed, an overall minus sign is included for all the terms of the ghost loop, as explained in section \ref{sec4}; this is the fermion-loop sign of the closed ghost line. The pure-gluon two-point results of the previous subsection illustrate the outcome of this sewing explicitly; the complete two-way two-loop kernels are provided in the ancillary file.

\subsection{General strategy for two-loop planar integrands}
The analysis of the two-point case contains all the structural elements of the general recursion, and it suggests the following strategy for an arbitrary $n$-point two-loop planar integrand:
\begin{enumerate}
    \item Based on \eqref{n-loop}, write down the formula for the two-loop planar integrands and identify all the loop kernels that are needed, i.e.\ the kernels $I^{\rm kernel}_{2,k}$ with $k=2,\ldots,n$ together with the one-loop kernels that enter their recursion.
    \item Compute the one-loop kernels directly, using the matrix formalism of section \ref{sec3} for the gluon parts and the analysis of section \ref{sec4} for the ghost parts. To construct the two-loop kernel $I_{2,m}^{\rm kernel}$ from one-loop kernels, only the following elements need to be considered:
    \ie
    &\frac{1}{2}I^{\rm gluon}_{1,i}\ (i\in\{2,3,\cdots,m+2\}), \\
    &(1-\frac{1}{2}\delta_{i,\frac{m}{2}+2})I^{\rm ghost}_{1,i}\ (i\in\{2,3,\cdots,\lfloor\frac{m}{2}\rfloor+2\})\\
    &\text{with}\ I^{\rm ghost}_{1,i}=I^{\rm kernel}_{1,i}-I^{\rm gluon}_{1,i}
    \fe
    Note that there is no additional graph factor $\frac{1}{\ell-r}$ in this reduction: for the two-loop kernels constructed by the recursion, $r=0$ in every term (the newly created loop shares more than one propagator with the largest loop only starting at three loops), and the factor $\frac{1}{\ell}=\frac12$ has already been taken into account in the coefficients above. In other words, due to the implicit symmetries of the construction, we do not need to process all $k$-way one-loop kernels independently.
    \item Assemble the reducible part of \eqref{n-loop} from the one-loop kernels and the one-loop currents $A^{(1)}$, computed as in \eqref{eq:oneloop-current}; assemble the irreducible part from the two-loop kernels of step 2.
    \item Construct each term of \eqref{n-loop} by replacing the kernel legs with the appropriate Berends-Giele or loop currents, and sum over the divisions of the cyclic set. This yields the complete two-loop planar integrand with all open legs gluonic.
\end{enumerate}

The two-way two-loop kernels that follow from this strategy have been computed both for the pure-gluon sector and for the complete YM theory with ghosts in the ancillary file, and they have been validated by comparing the resulting two-loop integrands with the Feynman-rule results. The same strategy applies, with only the obvious changes in the $k$-divisions, to the $n$-point two-loop integrands with $n>2$: the set of one-loop kernels that must be computed is fixed by the list of step 2, and the recursion then produces all two-loop kernels automatically.

\section{Conclusion and outlook}\label{sec6}
In this paper we have given a detailed account of the off-shell recursion for the planar loop integrands of the Yang-Mills theory, extending and completing the framework introduced in \cite{Tao:2025fch}. The main results are of three kinds.

First, we established the transfer-matrix formalism of the pure-gluon sector. We showed by an explicit induction that every comb component of a gluon multi-particle current is generated by a product of $2\times2$ block matrices, one per vertex of the comb, and that the one-loop kernel factorizes into a chain of such matrices together with a contact seed matrix. We spelled out the index and contraction conventions of the factorization and explained why it makes the sewing step local: a leg that is sewn into a kernel modifies only its neighboring transfer matrices, so that the re-evaluation of a term requires only a local substitution and a chain re-multiplication. This property is, in our view, the concrete structural advantage of the formalism over the direct Feynman-rule evaluation, and it is the property on which a future systematic derivation of higher-loop amplitude relations can be built.

Second, we completed the recursion by including the Faddeev-Popov ghosts. We derived the ghost multi-particle currents and their comb components from the equations of motion, fixed the Grassmann parity and sign conventions of the recursion once and for all, and translated them into definite sewing rules with orientation-dependent propagator signs. The complete one-loop kernels with $m=2,3,4$ ways, including the ghost contributions, have been checked against independent Feynman-rule computations; the two-way kernel is displayed in section \ref{sec5}, and the $m=3,4$ kernels are provided in the ancillary file. This closes the gap between the pure-gluon construction of \cite{Tao:2025fch} and the full YM theory, and it removes the main technical ambiguity of the ghost sector that was left open in the earlier discussion.

Third, we worked out the two-loop recursion in full detail. We enumerated the ordered sets and $k$-divisions of the two-point two-loop case, identified the one-loop kernels that are actually required, and explained, through the graph factors and the implicit symmetries of the recursion, why the remaining kernels need not be processed. The resulting strategy applies to general $n$-point two-loop planar integrands and reduces the computation of any two-loop kernel to a fixed, small list of one-loop kernels.

Several directions suggest themselves for future work. The matrix formalism of the pure-gluon sector is likely to be the right language for studying amplitude relations at higher loops: many relations, such as those obtained through differential operators that transmute all loops into scalar loops \cite{Cheung:2017ems,Chen:2023bji}, do not receive ghost contributions, so that the pure-gluon sector alone is sufficient and the nearest-neighbor structure of the transfer matrices turns a relation on a single leg into a relation on a fixed number of adjacent matrices. In addition, at the one-loop level, pure-gravity integrands are known to follow from the multi-particle currents \cite{Gomez:2024xec}; extending the present recursion to non-planar and gravity integrands at higher loops remains an open and important problem. We leave a systematic study of such directions for future work.

\section*{Acknowledgments}
YT thanks Chen Huang for the suggestions for the draft. YT is supported by Chinese NSFC funding NO. 124B2094.

\bibliographystyle{JHEP}
\bibliography{planarloop}
\clearpage
\appendix
\begin{widetext}
\section{Graph factors and the bare loop kernel}\label{app:graph}
In the loop kernel recursion, we need to define the \textit{graph factor} in order to avoid overcounting. For an $\ell$-loop Feynman diagram, a graph factor can be written as $g=S\times \frac{1}{\ell-r}$. Here, $S$ is the symmetry factor coming from the symmetry of the Feynman diagram, and $r$ is defined as below. We first define the largest loop to be the loop that is connected with all external legs and no other loop outside it. Then we consider all loops that are independent in this largest loop, i.e., the regions of different loops will not overlap. After choosing the largest loop, we will ignore the external legs and the 3-point vertices connected to them and regard the diagram just as a vacuum diagram. Then, each loop will have some propagators in common with the largest loop. If there is 1 common propagator, we assign the loop a value of 0; otherwise 1. This value is called the \textit{loop value}. Then $r$ is the sum of the values of all loops.

In addition, there is an extra contribution to the symmetry factor. For 2-way kernels, there is a symmetry under flipping the diagram. By the word ``flip", we mean ``turn upside down". This will lead to some overcounts when we do the recursion, since the two flipped configurations are produced by different sewings. Hence, we will regard the 2-way kernel and its flipped image as two different diagrams during the recursion, and each of them will get an extra symmetry factor of 1/2; if the two-way kernel and its flipped image coincide, no extra factor is needed. A concrete worked example of this counting in the bi-adjoint scalar theory, where all symmetry factors come from the two-way kernel parts of the diagrams, can be found in \cite{Tao:2025fch}; an elegant algebraic derivation of the same factors was given in \cite{Tao:2025pnt}.

The loop kernel without the symmetry factor $S$ in each diagram is called a \textit{bare loop kernel}. This concept is important when considering recursion, since when we go to higher-loop cases, the symmetry factor will change; hence, we only need to consider the symmetry factor in the last recursion. However, the factor $\frac{1}{\ell-r}$ here is to avoid overcounting in the following recursion; hence, we must consider it in every recursion. 

It is worth explaining the origin of the factor $\frac{1}{\ell-r}$ in the recursion, since this factor is sometimes introduced only as a rule. In the recursion of sections \ref{sec2} and \ref{sec4}, an $\ell$-loop kernel is generated by sewing a new loop onto an $(\ell-1)$-loop kernel. The newly generated loop always lies outside the previous kernel, and after the recursion is completed it is a subloop of the largest loop of the final diagram. A loop with value $0$ shares exactly one propagator with the largest loop; such a loop can be produced by the recursion, and it is produced exactly once for a given final diagram if the previous kernel is irreducible. A loop with value $1$ shares more than one propagator with the largest loop; such a loop cannot be generated by sewing onto an irreducible kernel in the way described, since that would require a reducible $(\ell-1)$-loop diagram that is not part of the kernel. The recursion therefore generates each irreducible $\ell$-loop diagram once for every admissible choice of the largest loop and of the loop values that sum to $r$, and the factor $(\ell-r)^{-1}$ cancels this overcounting. In the two-loop case all diagrams have $r=0$, because every independent loop shares exactly one propagator with the largest loop; this is why the factor $\frac{1}{\ell-r}$ never appears explicitly in section \ref{sec5}.

The separation of the two types of factors---the topological factor $\frac{1}{\ell-r}$ and the diagrammatic symmetry factor $S$---is what makes the graph-factor bookkeeping systematic: $S$ is fixed only in the last recursion step, when the final kernel is assembled from the bare kernel, while $\frac{1}{\ell-r}$ must be included at every step because it prevents overcounting between different recursion sequences. Both statements apply unchanged when ghost lines are included, since the two factors depend only on the loop topology and not on the particle content of the lines.

\end{widetext}

\end{document}